\documentclass[aps,prl,nofootinbib,twocolumn,letterpaper]{revtex4}

\usepackage{graphicx, color}
\usepackage{bm}
\usepackage{amsmath,amssymb,amsfonts}

\newcommand{\secret}[1]{}

\begin{document}
\thispagestyle{empty}

\noindent \textbf{Comment on ``Testing Planck-Scale Gravity with Accelerators''}

In recent paper \cite{Gharibyan:2012gp}, Gharibyan proposed a method of testing ``quantum gravity'' effects (such as energy-dependent vacuum refractive index and vacuum birefringence) with accelerators. The essence of the method was the experimental study of modifications of the photon's dispersion relation manifesting itself in a shift of the Compton edge (maximal energy of scattered photons) in a high-energy Compton scattering. Even though the idea is interesting on its own, the method and conclusions
are wrong, since, e.g., no effect on the electron's dispersion relation was considered.

\begin{figure}[t]
\centering
\includegraphics[width=8cm]{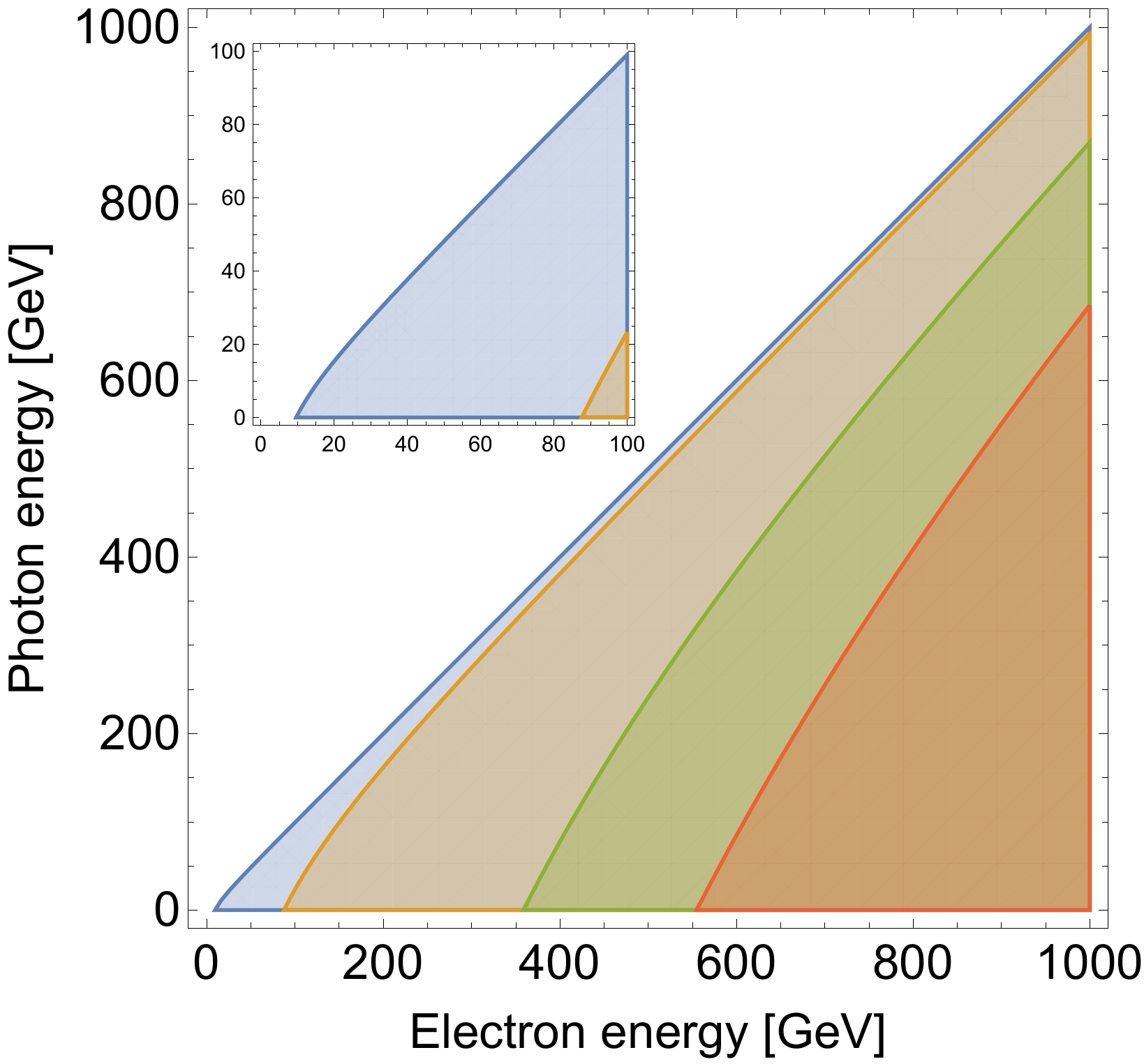}
\caption{\label{fig1}
Regions of allowed energies for an electron and a photon in the vacuum Cherenkov radiation at $n-1 = 1.4\cdot10^{-9}$, $1.7\cdot10^{-11}$, $10^{-12}$, $4.1\cdot10^{-13}$ (from left to the right).}
\end{figure}

From the phenomenological point of view, this assumption would lead to the vacuum Cherenkov radiation, $\mathrm{e}^- \rightarrow \mathrm{e}^- \gamma$. 
While such a process is kinematically forbidden for a standard dispersion relation of a photon of momentum $\textbf{k}$ and energy $\omega$, $\textbf{k}^2 = \omega^2$, it becomes possible for $\textbf{k}^2 = n^2\,\omega^2$, $n>1$, considered in the paper. If energy of the initial electron is $\mathcal{E}$, and the energy of the emitted photon is $\omega$, then, from the energy-momentum conservation, the condition on the emission is
\begin{align}
-1 \,\le\, \displaystyle\frac{\mathcal{E} - \omega n}{n \sqrt{\mathcal{(E-\omega})^2 - m_e^2}} \,\le 1\,,\quad \omega < \mathcal{E}\,.
\end{align}
Allowed regions of the process are demonstrated in Fig.~\ref{fig1}. One can see that $n-1 = 1.7\cdot 10^{-11}$ and $n-1 = 4.1\cdot 10^{-13}$ (proposed in Ref.~\cite{Gharibyan:2012gp}) would lead to the vacuum Cherenkov radiation at energy scales already explored experimentally. No such radiation was observed and the claimed values of $n-1$ were already excluded in accelerator experiments several years prior to the publication of Ref.~\cite{Gharibyan:2012gp} (see, e.g., Refs.~\cite{Hohensee:2009zk, Altschul:2009xh}). Also, when approaching the value $n-1 \sim 10^{-13}$ or when considering birefringence, additional Lorentz-violating parameters modifying dispersion relations must be considered, see Ref.~\cite{Hohensee:2009zk} and Refs. therein.

From the theoretical point of view, it is not clear, why considered ``quantum'' effects would not modify the electron's dispersion relation even if this happens already on the classical level -- within the General Relativity (GR), for an electron of the momentum $\textbf{p}$ and energy $\mathcal{E}$ it will be given by (see Refs.~\cite{AmelinoCamelia:2008qg, Kalaydzhyan})
\begin{align}
\textbf{p}^2 = \mathcal E^2 n^2 - m_e^2 /n^2\,.
\end{align}
In the high energy limit, $\mathcal{E} \gg m_e$, for the absolute value, $p = |\textbf{p}|$ this gives
\begin{align}
\frac{p}{\mathcal{E}} \simeq n - \frac{m_e^2}{2 \mathcal{E}^2}\,,\label{ne}
\end{align}
which should be used as an \textit{a priori} hypothesis in order to reproduce the classical limit, since it leaves the Compton edge and other quantities unaffected by gravity. One can check that small corrections to the right hand side of (\ref{ne}) and to $n$ are interchangeable and, hence, only their combination is measurable in a type of experiments proposed in Ref.~\cite{Gharibyan:2012gp}, see also comments in Ref.~\cite{Hohensee:2009zk}.

In addition, the quoted above numbers for the refractivity are too large to be in the domain of the quantum gravity effects (compare to Ref.~\cite{AmelinoCamelia:2008qg}). For instance, if the electron was not attracted to the Earth at all,
 this would correspond (in notations of Ref.~\cite{Gharibyan:2012gp}) to the effective refractive index $n-1 = 1.4\cdot 10^{-9}$, see Ref.~\cite{Kalaydzhyan}. If the electron was attracted to the Earth only by 1\% weaker, then $n-1 \sim 10^{-12}$, which falls in between the ``quantum'' refractivities quoted above. Clearly, both of the cases are also excluded on the basis of the absent vacuum Cherenkov radiation (see Fig.~\ref{fig1}).

A possible effect shifting the Compton edge in Ref.~\cite{Gharibyan:2012gp} might be the electromagnetic interaction of the electron with the beam and its images in the walls of the vacuum chamber (for polarization-independent results) -- this should be studied as
an additional systematic factor.

\secret{
From the experimental point of view,
 some of the important effects were not taken into account in Ref.~\cite{Gharibyan:2012gp} while interpreting the data. One of them is the electromagnetic interaction of the electron with the beam and its images in the walls of the vacuum chamber (for polarization-independent results) and polarization of the electrons in the beam due to the synchrotron radiation \cite{Barber:1992fc}
  (for the polarization-dependent effects).
}


\vspace{.3cm}
\noindent Tigran Kalaydzhyan

\begingroup\raggedright\leftskip=20pt
{\footnotesize
\noindent Department of Physics,
University of Illinois at Chicago,\\
845 W Taylor Street, Chicago, IL 60607, USA.}
\par\endgroup

\end{document}